\documentclass[twocolumn,amsmath,amssymb,aps,prl,superscriptaddress]{revtex4}
\usepackage{graphicx}
\usepackage{amsmath}
\usepackage{color}
\usepackage{array}
\usepackage{subeqnarray}
\usepackage{ulem}
\usepackage[english]{babel}
\usepackage{setspace}
\usepackage{hyperref}
\usepackage{cleveref}
\usepackage{bm}
\begin{document}
\title{Enhanced Dip Coating on a Soft Substrate}
\author{Vincent Bertin}
\email{v.l.bertin@utwente.nl}
\affiliation{Univ. Bordeaux, CNRS, LOMA, UMR 5798, 33405 Talence, France.}
\affiliation{UMR CNRS Gulliver 7083, ESPCI Paris, PSL Research University, 75005 Paris, France.}
\affiliation{Physics of Fluids Group, Faculty of Science and Technology, and Mesa+ Institute, University of Twente, 7500AE Enschede, The Netherlands.}
\author{Jacco H. Snoeijer}
\affiliation{Physics of Fluids Group, Faculty of Science and Technology, and Mesa+ Institute, University of Twente, 7500AE Enschede, The Netherlands.}
\author{Elie Rapha\"{e}l}
\affiliation{UMR CNRS Gulliver 7083, ESPCI Paris, PSL Research University, 75005 Paris, France.}
\author{Thomas Salez}
\affiliation{Univ. Bordeaux, CNRS, LOMA, UMR 5798, 33405 Talence, France.}
\date{\today}

\begin{abstract}

A solid, withdrawn from a liquid bath, entrains a thin liquid film. This simple process, first described by Landau, Levich and Derjaguin (LLD), is commonly observed in everyday life. It also plays a central role in liquid capture by animals, and is widely used for surface-coating purposes in industry. Motivated by the emerging interest in the mechanics of very soft materials, and in particular the resulting elastocapillary coupling, we develop a dip-coating model that accounts for the additional presence of a soft solid layer atop the rigid plate. The elastic response of this soft layer is described by a Winkler's foundation. Using a combination of numerical, scaling and asymptotic-matching methods, we find a new softness-dependent power-law regime for the thickness of entrained liquid at small capillary number, which corresponds to a modified physics at play in the dynamic meniscus. The crossover between this regime and the classical dip-coating one occurs when the substrate's deformation is comparable to the thickness of the entrained liquid film. 
\end{abstract}

\maketitle

\begin{figure}[t!]
\centering
\includegraphics[width=0.8\columnwidth]{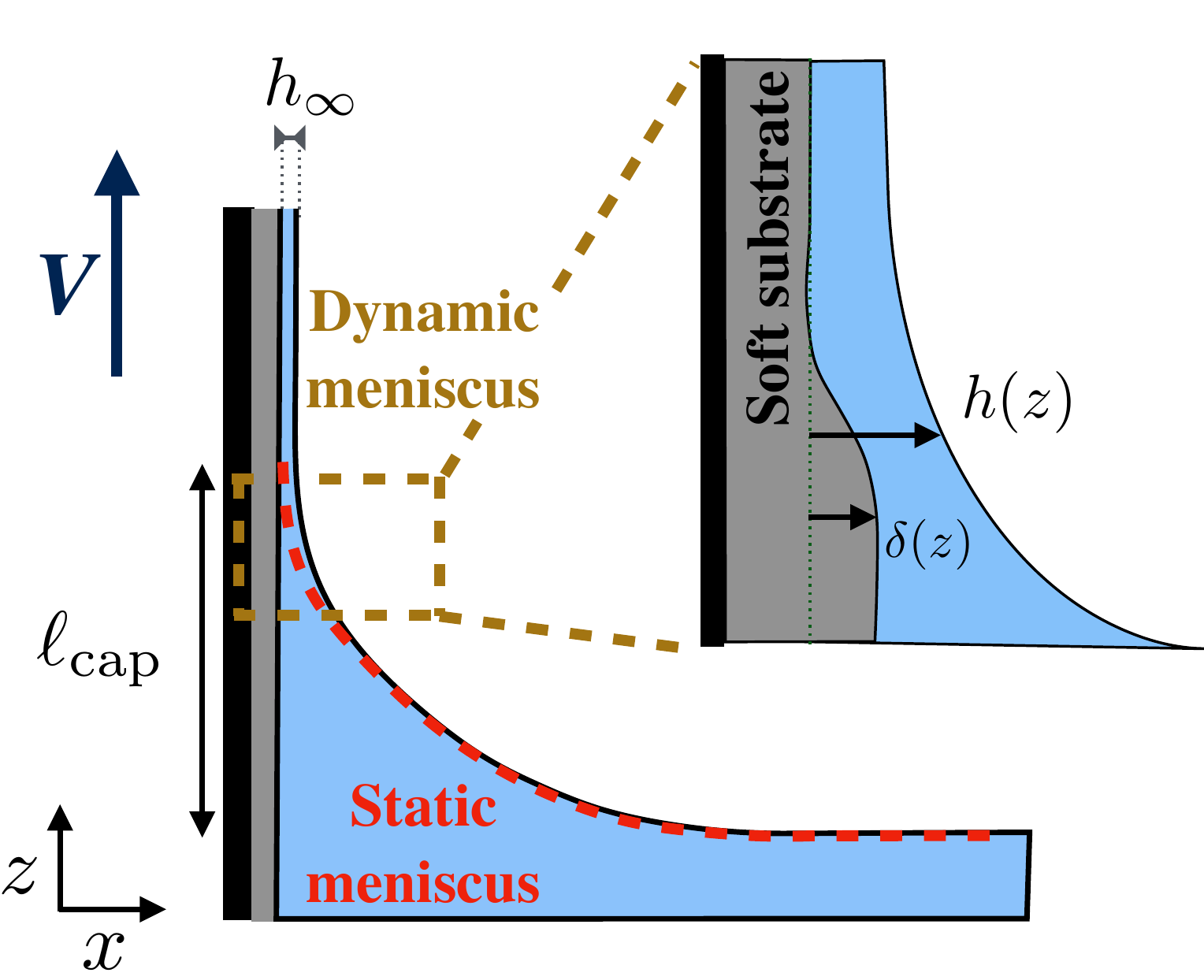}
\caption{Schematic of the soft dip-coating problem. The inset exhibits a zoom in the dynamic-meniscus zone where the flow is localized. The liquid-air interfacial profile is denoted $h(z)$ and its curvature induces a negative Laplace pressure that generates a deformation $\delta(z)$ of the soft substrate. }
\label{fig:schema}
\end{figure}

A solid object, withdrawn from a liquid bath, entrains a thin liquid film via viscous forces. Such a process is called \textit{dip coating} and is commonly used in industry for surface treatment with specific (\textit{e.g.} optical) properties~\cite{baumeister2004optical,brinker2013dip}. The central quantity of interest is the thickness $h_\infty$ (see Fig.~\ref{fig:schema}) of the entrained liquid film. Using asymptotic-matching methods, Landau, Levich~\cite{levich1942dragging} and Derjaguin~\cite{derjaguin1943thickness} were the first ones to calculate this thickness for a Newtonian liquid coating a rigid substrate. Over the last decades, the LLD description has been challenged in several ways~\cite{rio2017withdrawing}. The film thickness has been shown to drastically depend on fluid inertia~\cite{de1998gravity,jin2005drag}, the presence of surfactants at the liquid-air interface~\cite{shen2002fiber}, the non-Newtonian properties of the liquid~\cite{spiers1975free,ro1995viscoelastic,ashmore2008coating,smit2019stress,Marchand2020,datt2021thin}, or the roughness of the solid~\cite{krechetnikov2005experimental,seiwert2011coating}, to cite a few. 

Besides, a recent and growing interest was devoted to the  mechanics of soft materials (Young's modulus $E \sim \mathrm{kPa}$), with a plethora of applications towards micrometric and biomimetic systems. When wetted by droplets, such materials exhibit rich soft-wetting properties, as contact-line capillary forces (through \textit{e.g.} the liquid-air surface tension $\gamma$) are sufficient to deform them~\cite{bico2018elastocapillarity}. This generates \textit{e.g.} interfacial ridges~\cite{style2017elastocapillarity, andreotti2020statics}, which considerably change the spreading and motion of droplets in comparison with the case of rigid substrates~\cite{karpitschka2016liquid,hourlier2017role}. 

In a dip-coating process involving soft surfaces, the Laplace pressure is expected to induce elastic deformations that modify the flow and the entrained thickness (\textit{cf.} Fig.~\ref{fig:schema}). This situation is reminiscent of soft levelling~\cite{Rivetti2017}, and might have implications in liquid capture by animals~\cite{kim2011optimal,kim2012natural,Brau2016}, through \textit{e.g.} the softness and geometry of the 
tongue~\cite{Lechantre2021}. In this Letter, we thus study theoretically the influence of the elastic deformation of a soft substrate on the thickness of liquid entrained in dip coating. Mainly, a new, \textit{soft-LLD} regime is identified, and characterized using a similarity solution. 

A schematic of the system is shown in Fig.~\ref{fig:schema}. We consider a rigid substrate covered by a thin compressible linear-elastic layer, modeled as a Winkler's foundation~\cite{dillard2018review}. The ensemble is withdrawn with a velocity $V$ from a liquid reservoir of viscosity $\eta$ and density $\rho$. The problem is assumed to be invariant in the $y$ direction, and is solved using a LLD-like asymptotic-matching method~\cite{levich1942dragging,wilson1982drag}. We focus on the dynamic meniscus zone (see inset of Fig.~\ref{fig:schema}) and use the lubrication approximation to characterize the steady liquid-air interface profile $h(z)$. Gravitational drainage is neglected, so that the hydrodynamic pressure field is set by the Laplace pressure. The key element introduced in this work is that the hydrodynamic pressure induces a normal elastic deformation $\delta(z)$ of the soft layer. The thickness of the liquid layer is thus given by $h-\delta$, which modifies the thin-film equation~\cite{oron1997long} to the form
\begin{equation}
\label{eq:thinfilm}
\frac{\gamma}{3\eta} \left[h(z)-\delta(z)\right]^3 h'''(z) + V \left[h(z)-\delta(z)\right] = V h_\infty\ ,
\end{equation}
where a prime denotes one spatial derivative with respect to $z$, and $V h_\infty$ is the flow rate (per unit length). Far from the bath, the liquid-film thickness reaches $h_\infty$, such that 
\begin{equation}\label{eq:bc_film}
h(z\to \infty) = h_\infty\, .
\end{equation}
Following the standard procedure~\cite{wilson1982drag,de2013capillarity}, matching to the static meniscus is achieved via the boundary condition
\begin{equation}
\label{eq:matching_condition}
h''(z\to -\infty) = \frac{\sqrt{2}}{\ell_\mathrm{cap}}\ ,
\end{equation}
where $\ell_\mathrm{cap}=\sqrt{\gamma/(\rho g)}$ is the capillary length. 

To close the problem, we need to specify the deformation $\delta$. As a minimal description of the mechanical response of the elastic layer, we use the Winkler's foundation~\cite{dillard2018review}, that is valid for thin-enough compressible materials under small deformations~\cite{chandler2020validity}. Essentially, the soft layer is described as a mattress of independent springs. Thus, the normal deformation is simply proportional to the local pressure, as
\begin{equation}
\label{eq:winkler}
\delta(z) = \frac{t \gamma}{E^*} h''(z)\ ,
\end{equation} 
with the layer thickness $t$ and the effective modulus $E^* = \frac{E(1-\nu)}{(1+\nu)(1-2\nu)}$, where $\nu$  denotes the Poisson ratio ($\nu\neq1/2$). The length scale $\ell_\mathrm{ec}=\sqrt{t \gamma/E^*}$ is the relevant elastocapillary length for the current geometry and elastic response~\cite{bico2018elastocapillarity}. Inserting Eq.~\eqref{eq:winkler} into Eq.~\eqref{eq:thinfilm}, we find a closed differential equation
\begin{equation}
\label{eq:softLLD}
(h - \ell_\mathrm{ec}^2 h'')^3 h''' = 3\mathrm{Ca} \, \left(h_\infty - h + \ell_\mathrm{ec}^2 h''\right)\ ,
\end{equation}
where $\mathrm{Ca}=\eta V/\gamma$ is the capillary number. The equation contains $\ell_{\rm ec}$, while the matching condition (see Eq.~\eqref{eq:matching_condition}) involves $\ell_{\rm cap}$. Therefore, besides ${\rm Ca}$, the problem contains another dimensionless softness parameter $\mathcal{L} = (\ell_\mathrm{ec}/\ell_\mathrm{cap})^2 = \rho g t/E^*$, which characterizes the relative importance of softness in the problem.

We numerically solve Eq.~\eqref{eq:softLLD} using a $4^{\mathrm{th}}$-order Runge-Kutta scheme, where the boundary condition of Eq.~\eqref{eq:bc_film} is imposed via the solution of the linearized version of Eq.~\eqref{eq:softLLD}~\cite{numerical}. The numerical solution behaves as $h \sim z^2$ as $z\to -\infty$, and we adapt the value of $h_\infty$ via a shooting algorithm to achieve the curvature-matching condition of Eq.~\eqref{eq:matching_condition}.
\begin{figure}[t!]
\centering
\includegraphics{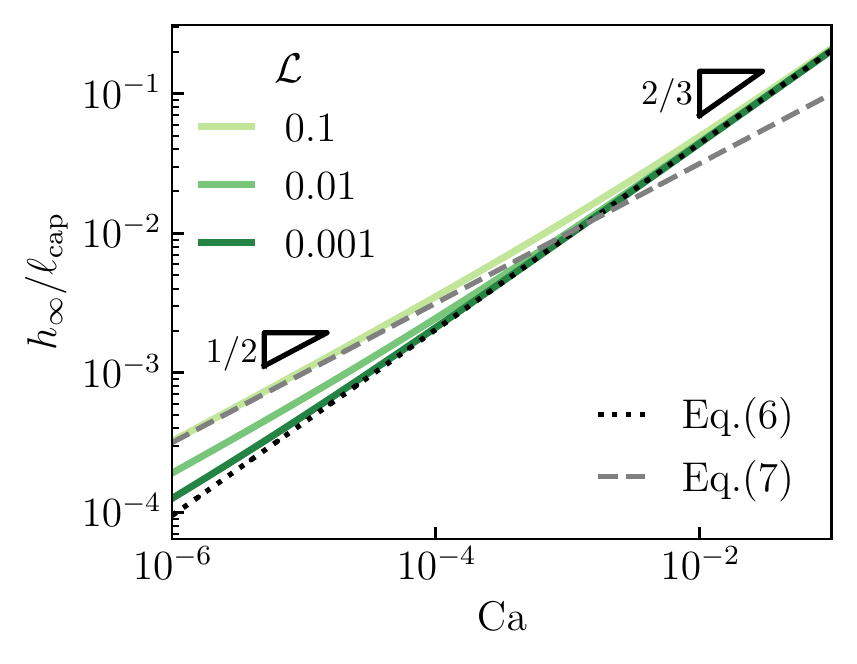}
\caption{Normalized thickness of the entrained liquid film versus capillary number, for different values of the dimensionless softness parameter $\mathcal{L} = \rho g t / E^*$, as obtained from the numerical integration of Eq.~\eqref{eq:softLLD}. The black dotted line represents the classical LLD law of Eq.~\eqref{eq:Landau-Levich} and the gray dashed line shows the soft-LLD law of Eq.~\eqref{eq:soft_thickness}. The slope triangles indicate the power-law exponents of Eqs.~(\ref{eq:Landau-Levich})~and~(\ref{eq:soft_thickness}).}
\label{fig:thickness}
\end{figure}

Figure~\ref{fig:thickness} reports the normalized thickness of the entrained liquid film as a function of the capillary number, for three different dimensionless softness parameters. Two distinct scaling regimes can be observed. At large $\mathrm{Ca}$, we recover the classical LLD power law of the rigid case~\cite{levich1942dragging}
\begin{equation}
\label{eq:Landau-Levich}
h_\infty \approx 0.946 \, \ell_\mathrm{cap} \, \mathrm{Ca}^{2/3}\ .
\end{equation}
At finite values of the softness parameter $\mathcal L$, however, one finds that the result deviates from Eq.~\eqref{eq:Landau-Levich} at small $\mathrm{Ca}$. The larger $\mathcal L$, the stronger the departure from the classical scaling. At small dip-coating velocities, we find a novel, soft-LLD power-law regime for which the thickness of the entrained liquid film given by
\begin{equation}
\label{eq:soft_thickness}
h_\infty = \frac{2}{3\sqrt[4]{2}} \sqrt{\ell_\mathrm{cap}\ell_\mathrm{ec}} \, \mathrm{Ca}^{1/2}\ .
\end{equation} 
The determination of the scaling and prefactor of Eq.~\eqref{eq:soft_thickness} will be discussed below. Our central result is thus that, at small-enough velocity, the wall softness enhances the dip-coating efficiency with respect to the classical LLD scenario. 

\begin{figure}[t!]
\centering
\includegraphics[width=\columnwidth]{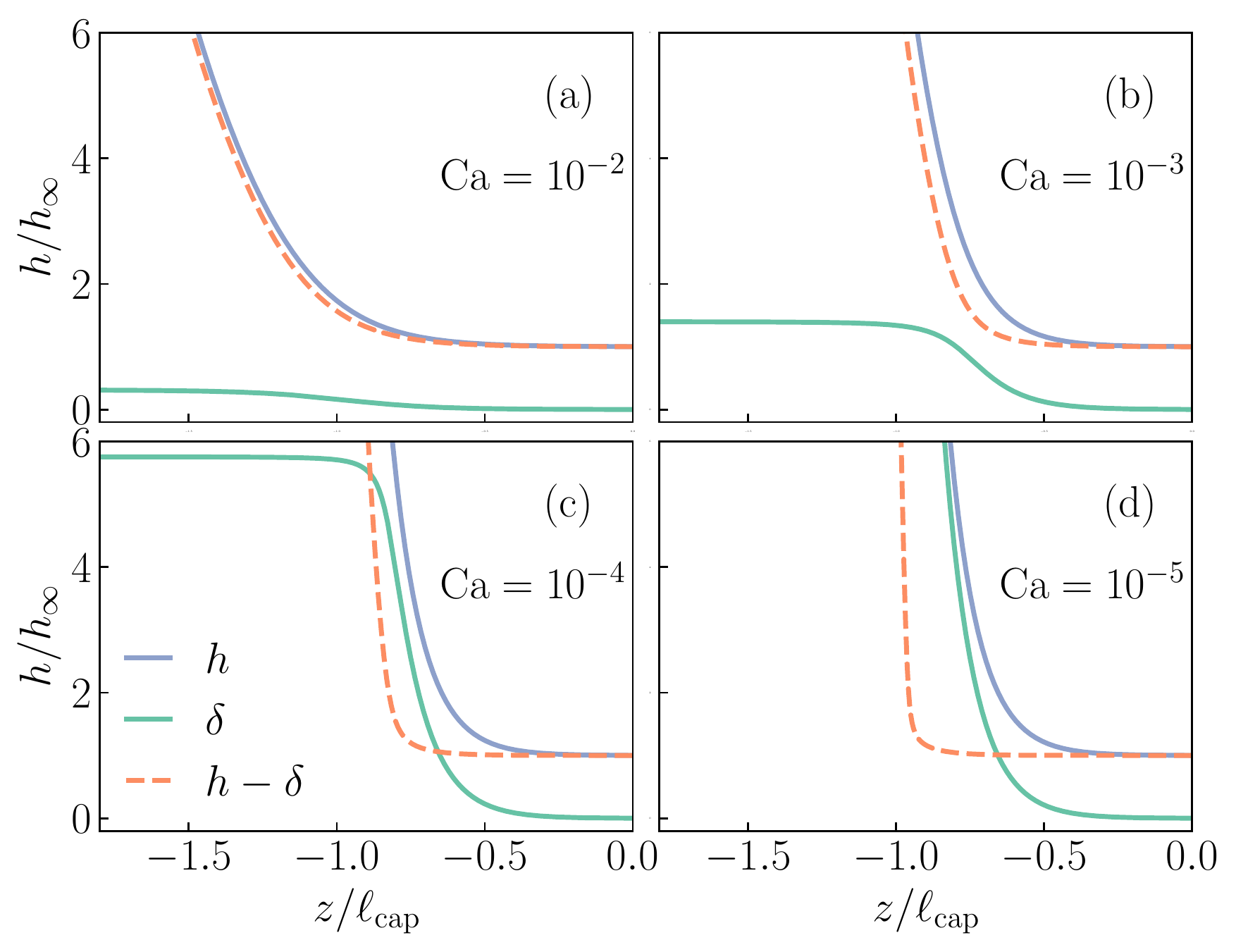}
\caption{Profiles of the liquid-air interface (blue) and the normal deformation of the elastic layer (green) normalized by the thickness of the entrained liquid film, as functions of the vertical position normalized by the capillary length, as obtained from the numerical integration of Eq.~\eqref{eq:softLLD}. The orange dashed lines display the normalized thickness profiles of the liquid layer. The dimensionless softness parameter is set to $\mathcal{L}=0.01$. The capillary numbers are $\mathrm{Ca} = 10^{-2}, 10^{-3}, 10^{-4}$ and $10^{-5}$, in (a)-(d) respectively. }
\label{fig:profiles}
\end{figure}

The emergence of a soft-LLD regime at low velocity can be understood by comparing the typical elastic deformation to the thickness of the entrained liquid film. As $z\to -\infty$, the normal deformation of the soft layer, that is proportional to the curvature of the liquid-air interface in the Winkler's model, reaches 
\begin{equation}
\label{eq:normal_deformation_limit}
\delta(z\to -\infty) = \sqrt{2} \, \ell_\mathrm{ec}^2/\ell_\mathrm{cap}\ ,
\end{equation} 
as obtained from injecting Eq.~\eqref{eq:matching_condition} into Eq.~\eqref{eq:winkler}. Interestingly, Eq.~\eqref{eq:normal_deformation_limit} does not involve the velocity. Therefore, the relative magnitude of the elastic deformation versus the thickness of the entrained liquid film increases with decreasing ${\rm Ca}$. This explains why the soft regime emerges at small velocity. Estimating the rigid-to-soft crossover to take place when the normal elastic deformation and the thickness of the entrained liquid film are of the same order (\textit{i.e.} $\delta(z\to-\infty) \sim \ell_{\rm cap}{\rm Ca}^{2/3}$), we obtain a critical capillary number ${\rm Ca}^* \sim \mathcal L^{3/2}$, which is exactly the scaling obtained by balancing Eqs.~\eqref{eq:Landau-Levich} and \eqref{eq:soft_thickness}. To verify the above scenario, we show in Fig.~\ref{fig:profiles} both the liquid-air interface profile and the normal deformation profile of the elastic layer versus the vertical position, for a situation where $\mathrm{Ca}^* =10^{-3}$. For $\mathrm{Ca}<\mathrm{Ca}^*$, the normal elastic deformation is found to be larger than the thickness of the entrained liquid film (and vice versa), thus confirming the above criterion. 

\begin{figure}[t!]
\centering
\includegraphics[width=0.8\columnwidth]{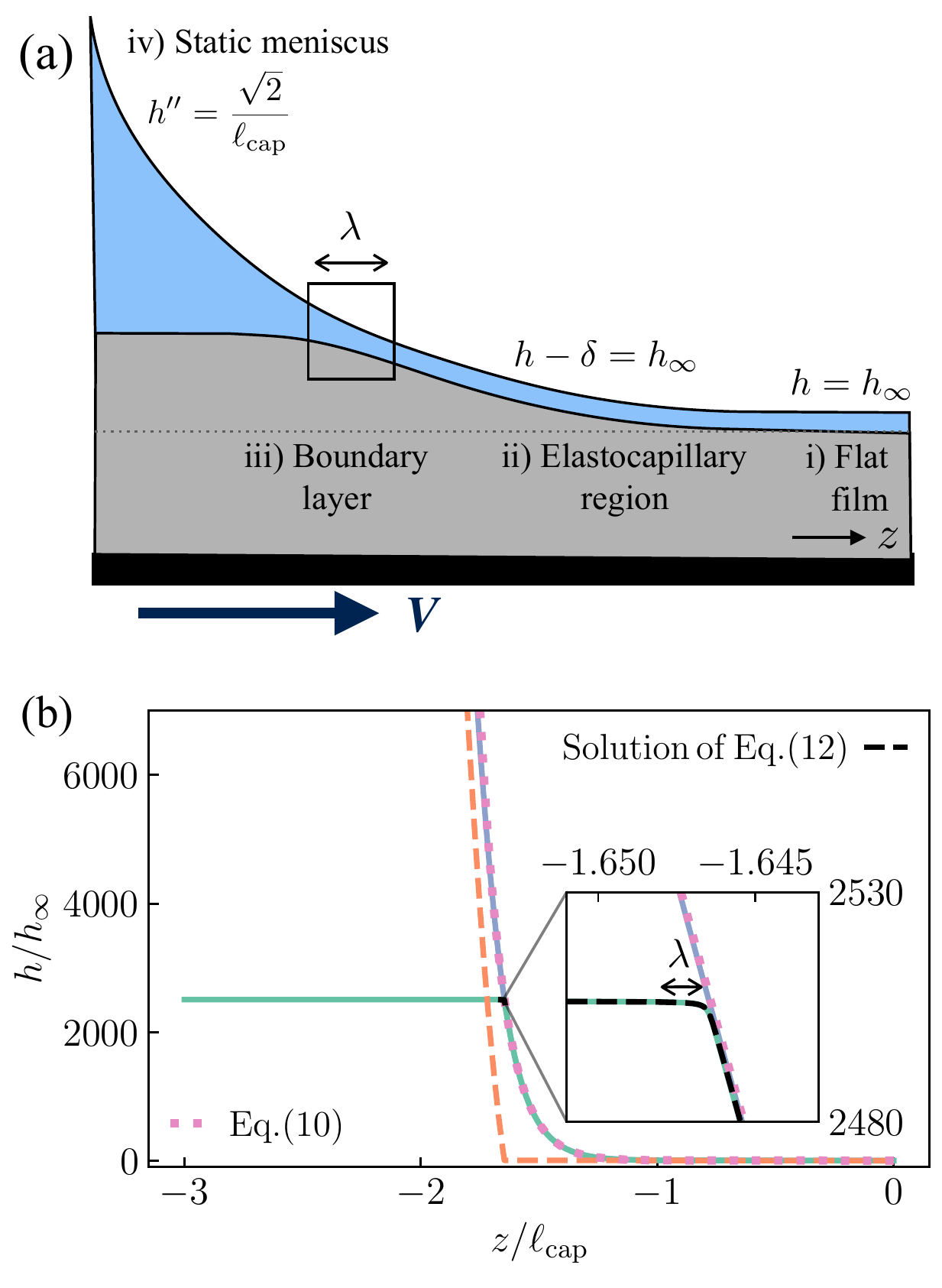}
\caption{(a) Schematic showing the four regions of interest: i) at large $z$, the entrained liquid film is flat and the elastic layer is not deformed; ii) reducing $z$, one enters the elastocapillary region, with a constant fluid thickness and a progressively-deformed elastic layer; iii) a dynamic meniscus (``boundary layer") then connects the elastocapillary region to the static meniscus; iv) the static meniscus. Note that the schematic is reoriented with respect to Fig.~\ref{fig:schema}. (b) Profiles of the liquid-air interface (blue) and the normal deformation of the elastic layer (green) normalized by the thickness of the entrained liquid film, as functions of the vertical position rescaled by the capillary length, as obtained from the numerical integration of Eq.~\eqref{eq:softLLD} with $\mathrm{Ca}= 10^{-9}$ and $\mathcal{L} = 0.01$. The orange dashed line displays the normalized thickness profile of the liquid layer. The dotted pink line corresponds to the outer elastocapillary solution given by Eq.~\eqref{eq:soft_zone}. The inset shows a zoom of the boundary layer, where the black dashed line displays the boundary-layer solution of Eq.~\eqref{eq:boundary_layer}. }
\label{fig:profile_soft}
\end{figure}

We now turn to a detailed analysis of the soft LLD regime, including the derivation of Eq.~\eqref{eq:soft_thickness}. Figure~\ref{fig:profile_soft}(a) depicts the typical problem structure in the soft regime, \textit{i.e.} ${\rm Ca} \ll {\rm Ca}^*$, through four main regions. In contrast to the rigid case, the flat-film region (i) does not immediately connect to the dynamic meniscus/boundary layer (iii): instead, one observes an intermediate elastocapillary region (ii) for which the elastic layer deforms significantly while the liquid-film thickness $h-\delta$ remains approximately constant and equal to $h_\infty$ (see also Figs.~\ref{fig:profiles}(c) and (d)). This fundamentally changes the structure of the boundary layer, and the subsequent matching to the static meniscus (iv). 

We describe the elastocapillary region by assuming that it is essentially static, \textit{i.e.} devoid of hydrodynamics, owing to the nearly-constant liquid-film thickness. Inserting $h-\delta =h_\infty$ in Eq.~\eqref{eq:winkler} and solving the obtained differential equation, we get the liquid-air interface profile in the elastocapillary region 
\begin{equation}\label{eq:static_elastocapillary}
h_{\rm ec}(z) \simeq h_\infty + A\ \textrm{e}^{-z/\ell_{\rm ec}} \ ,
\end{equation}
where $A$ is an integration constant. Note that this solution corresponds to nullifying the right-hand-side of Eq.~\eqref{eq:softLLD}, \textit{i.e.} by removing hydrodynamic effects.

Next, we need to connect the static elastocapillary solution (ii) to the static meniscus (iv). The latter exhibits a finite curvature $\sqrt{2}/\ell_{\rm cap}$, so that the exponentially-growing curvature of the elastocapillary solution in Eq.~\eqref{eq:static_elastocapillary} with decreasing $z$ must saturate. This saturation is achieved via a hydrodynamic boundary layer (iii). Introducing the location $z_{\textrm{c}}$ of the boundary layer, we first rewrite the outer elastocapillary solution as
\begin{equation}
\label{eq:soft_zone}
h_{\rm ec}(z) \simeq h_\infty + \sqrt{2}\frac{ \ell_{\rm ec}^2}{\ell_{\rm cap}}\textrm{e}^{(z_{\textrm{c}}-z)/\ell_{\rm ec}}\ .
\end{equation}
This elastocapillary solution perfectly describes the numerical liquid-air interface profile for $z > z_\mathrm{c}$ (see Fig.~\ref{fig:profile_soft}(b)). Below $z_\mathrm{c}$, the elastic deformation $\delta$ saturates to its limiting value of Eq.~\eqref{eq:normal_deformation_limit}. The inset of Fig.~\ref{fig:profile_soft}(b) provides a zoom in the boundary layer where the elastic deformation smoothly approaches its saturation value.

To characterize the boundary layer, we consider the vicinity of $z_{\textrm{c}}$, and define the similarity variable $\xi = (z-z_{\textrm{c}})/\lambda$, where $\lambda$ is the unknown boundary-layer width (see Fig.~\ref{fig:profile_soft} (a)). We make the following Ansatz
\begin{equation}
\label{eq:similarity_ansatz}
h(z) = h_\infty + \sqrt{2}\frac{ \ell_{\rm ec}^2}{\ell_{\rm cap}} \left( 1 - Z +\frac{1}{2} Z^2\right) + B \mathcal H (\xi)\ .
\end{equation}
This expression contains the second-order expansion of $h_{\rm ec}(z)$ expressed with the variable $Z=(z-z_{\textrm{c}}) /\ell_{\rm ec}$, while the boundary layer is described by a self-similar function $\mathcal H(\xi)$ and a constant $B$. The self-similar function must ensure the saturation of the curvature, and for that reason we define the natural auxilliary function $\mathcal K = \mathcal H''$. The boundary condition in Eq.~\eqref{eq:matching_condition} imposes $\mathcal K(\xi\to -\infty)= 0$. On the other side, matching to the third order of the expansion of $h_{\rm ec}(z)$ requires that $\mathcal{K}(\xi\to \infty) = -\sqrt{2}\xi\lambda^3/(B\ell_{\rm ec}\ell_{\rm cap})$, which after setting $B = \sqrt{2} \lambda^3/(\ell_{\rm ec}\ell_{\rm cap})$ reduces to $\mathcal K(\xi \to \infty)=-\xi$.
Inserting Eq.~\eqref{eq:similarity_ansatz} in Eq.~\eqref{eq:softLLD}, we obtain at leading order in $\lambda$

\begin{equation}\label{eq:boundary_layer}
\mathcal K' = - 3 \frac{\mathcal K+\xi}{(\mathcal K + \xi  - H_\infty)^3 }\ ,
\end{equation}
where we set $\lambda = 2^{-3/4} {\rm Ca}^{1/2} \ell_{\rm cap}^{3/2}\ell_{\rm ec}^{-1/2}$ to remove the capillary number from the problem, and we introduce $h_\infty = \sqrt{2}\ell_{\rm ec}\lambda H_\infty/\ell_{\rm cap}$, with $H_{\infty}$ a numerical prefactor. These relations lead to
\begin{equation}
h_\infty = 2^{-1/4} H_\infty \sqrt{ \ell_{\rm ec} \ell_{\rm cap}} {\rm Ca}^{1/2}\ ,
\end{equation}
where we recover the scaling of Eq.~\eqref{eq:soft_thickness}.

The remaining task is to solve Eq.~\eqref{eq:boundary_layer} subjected to the boundary conditions, which will select the value of $H_\infty$. Towards the static meniscus, \textit{i.e.} as $\xi \to -\infty$, Eq.~\eqref{eq:boundary_layer} has an asymptotic solution of the form $\mathcal K(\xi)\simeq 3/\xi$ which does not depend on $H_\infty$. Besides, towards the elastocapillary region, we expect the asymptotic behavior of the boundary layer to be of the form $\mathcal{K} \simeq -\xi + C + \mathcal K_1(\xi) $ where $\mathcal K_1(\xi)$ is a function vanishing at $\xi \to \infty$. Here, the asymptotic solution does depend on $H_\infty$ as $C$ must satisfy $3C/(C-H_\infty)^3=1$. Performing a linear-stability analysis, we find that $H_\infty = 2/3$ (and thus $C = -1/3$) is the only value that ensures an algebraic decay of $\mathcal{K}_1$ at large $\xi$, as required for the matching to the elastocapillary region~\cite{linear_stability_analysis,hinch_1991}.

The solution of Eq.~\eqref{eq:boundary_layer} with $H_\infty=2/3$ is plotted in the inset of Fig.~\ref{fig:profile_soft}(b), offering a perfect description of the elastic deformation inside the boundary layer. More importantly, $H_\infty=2/3$ provides the sought-after prefactor present in Eq.~\eqref{eq:soft_thickness}, which is in perfect agreement with direct numerical integration of Eq.~\eqref{eq:softLLD} (see Fig.~\ref{fig:thickness}).

We conclude here by some rough estimates towards practical relevance of the soft-LLD scenario exhibited in this Letter. In the case of a substrate coated with a thick elastic layer, the deformation does not depend anymore on the layer thickness $t$, and the relevant elastocapillary number becomes $\gamma/E$~\cite{bico2018elastocapillarity,andreotti2020statics}. If the scenario identified in the current work through a Winkler's foundation remains valid for other elastic responses, the crossover to the soft-LLD regime should occur for thick elastic materials at a critical capillary number $\mathrm{Ca}^*\sim (\frac{\gamma}{E\ell_\mathrm{cap}})^{3/2}$. Using typical values for soft gels, \textit{i.e.} $\ell_\mathrm{cap} \approx 1$ mm and $\gamma/E \approx 10\, \mu$m, we find a critical capillary number on the order of $10^{-3}$, which is in the accessible range experimentally~\cite{rio2017withdrawing}.   

As a perspective, extensions of the present model to other forms of elastic response and comparisons to experiments would be interesting. In addition, viscoelastic properties of the soft solid may affect the results~\cite{lhermerout2016moving}. Lastly, the displacement of a liquid meniscus on a solid occurs in various other situations~\cite{cantat2013liquid}, such as the motion of confined bubbles in a channel~\cite{bretherton1961motion}, or the spreading of a droplet~\cite{tanner1979spreading,lister2013viscous}. These problems also involve LLD-like solutions. Hence, it would be interesting to revisit them with soft boundaries using the present soft-LLD theory~\cite{charitatos2020thin}.

\section{Acknowledgements}
We thank F. Boulogne, M. Marchand, C. Poulard, F. Restagno and E. Rio for interesting discussions. This work is supported by the Agence Nationale de la Recherche (ANR) under the \textit{EMetBrown}  (ANR-21-ERCC-0010) and \textit{Softer} (ANR-21-CE06-0029) grants, and by the NWO through VICI Grant No. 680-47-632. We also thank the Soft Matter Collaborative Research Unit, Frontier Research Center for Advanced Material and Life Science, Faculty of Advanced Life Science at Hokkaido University, Sapporo, Japan.


\begin{thebibliography}{40}
\bibitem[{num()}]{numerical}
\bibinfo{note}{Assuming a flat film at $z\to \infty$, we linearize
  Eq.~\eqref{eq:softLLD} at large $z$ through $h(z) = h_\infty + \delta h(z)$,
  where $\delta h(z) \ll h_\infty$. This gives $\delta h'''(z) =
  \frac{3\mathrm{Ca}}{h_\infty^3}\left[\ell_\mathrm{ec}^2 \delta h''(z)-\delta
  h(z)\right]$. The solution of the latter, that is compatible with a flat film
  at $z \to \infty$, takes the form $\delta h(z) \propto \exp(rz)$, where $r$
  is the root of the polynomial equation $r^3 +
  \frac{3\mathrm{Ca}}{h_\infty^3}\left[-\ell_\mathrm{ec}^2r^2 + 1\right] = 0$,
  with a negative real part.}
\bibitem[{lin()}]{linear_stability_analysis}
\bibinfo{note}{Assuming $\mathcal{K}_1$ to be small, the linearization of
  Eq.~\eqref{eq:boundary_layer} gives $\mathcal{K}_1' = 3\mathcal{K}_1(2C +
  H_\infty)/(C-H_\infty)^4$. Thus, stable solutions at $\xi \to \infty$ only
  exist for $2C+H_\infty\leq 0$ and an algebraic decay of $\mathcal{K}_1$ is
  obtained $H_\infty = -2C$. Combining with the equation for $C$ (in the text),
  we get the numerical prefactor $H_\infty = -2C = 2/3$.}
\end{thebibliography}
\end{document}